\documentclass{article} 
\usepackage{graphicx} 
\usepackage{hyperref, amssymb, amsmath}
\usepackage[capitalise]{cleveref}
\usepackage{xcolor}
\usepackage[]{geometry}
\numberwithin{figure}{section}

\usepackage{tikz}
\usetikzlibrary{
intersections, arrows.meta,
automata,er,calc,
backgrounds,
mindmap,folding,
patterns,
decorations.markings,
fit,
shapes,matrix,
positioning,
shapes.geometric,
arrows,through
}

\title{Delayed loss of stability of periodic travelling waves: insights from the analysis of essential spectra}
\author{Lukas Eigentler$^{1}$, Mattia Sensi$^{2}$\\[1em]
$^1${\footnotesize Evolutionary Biology Department, Universit\"at Bielefeld, Konsequenz 45, 33615 Bielefeld, Germany}\\
$^2${\footnotesize Department of Mathematical Sciences ``G. L. Lagrange'', Politecnico di Torino,}\\ {\footnotesize Corso Duca degli Abruzzi 24, 10129 Torino Italy}}
\date{\today}

\usepackage[backend=bibtex, bibstyle=authoryear, citestyle=authoryear, giveninits=true, maxcitenames=2, sorting=nyt, sortcites=true, maxbibnames=99, isbn=false, url=false, doi=true, date=year, eprint=true, dashed=false, uniquelist=false, uniquename=false]{biblatex}

\newcommand{\dif}{\ensuremath{\operatorname{d}\!}}
\newcommand{\R}{\ensuremath{\mathbb{R}}}
\newcommand{\N}{\ensuremath{\mathbb{N}}}
\newcommand{\C}{\ensuremath{\mathbb{C}}}
\newcommand{\PTW}{\ensuremath{\operatorname{PTW}}}
\newcommand{\tstab}{t_{\operatorname{stab}}}

\newcommand{\tdelay}{t_{\operatorname{delay}}}
\newcommand{\Ajump}{A_{\operatorname{change}}}
\newcommand{\deltajump}{\delta_{\operatorname{change}}}
\newcommand{\Astab}{A_{\operatorname{stab}}}
\newcommand{\deltastab}{\delta_{\operatorname{stab}}}

\newcommand{\Atarget}{A_{\operatorname{target}}}
\newcommand{\deltatarget}{\delta_{\operatorname{target}}}
\newcommand{\mujump}{\mubar_{\operatorname{change}}}
\newcommand{\mubar}{\overline{\mu}}

\begin{document}
	
	\maketitle
\begin{abstract}
    Periodic travelling waves (PTW) are a common solution type of partial differential equations. Such models exhibit multistability of PTWs, typically visualised through the Busse balloon, and parameter changes typically lead to a cascade of wavelength changes through the Busse balloon. In the past, the stability boundaries of the Busse balloon have been used to predict such wavelength changes. Here, motivated by anecdotal evidence from previous work, we provide compelling evidence that the Busse balloon provides insufficient information to predict wavelength changes due to a delayed loss of stability phenomenon. Using two different reaction-advection-diffusion systems, we relate the delay that occurs between the crossing of a stability boundary in the Busse balloon and the occurrence of a wavelength change to features of the essential spectrum of the destabilised PTW. This leads to a predictive framework that can estimate the order of magnitude of such a time delay, which provides a novel ``early warning sign'' for pattern destabilization. We illustrate the implementation of the predictive framework to predict under what conditions a wavelength change of a PTW occurs. 
\end{abstract}	
	\section{Introduction}

Partial Differential Equations (PDEs) are ubiquitous in the field of mathematical modelling of spatio-temporal natural phenomena. In particular, an ever-growing number of researchers devote their attention to pattern formation in various PDE models to understand self-organisation in a wide range of fields, ranging from ecology \parencite{Rietkerk2008}, to cell biology \parencite{Dzianach2019}, to solar dynamics \parencite{Pontin2022}, and more.

This work focuses on a specific type of spatio-temporal pattern in PDE systems: periodic travelling waves (PTWs), sometimes referred to as wavetrains or plane waves. PTWs describe spatio-temporal patterns which are periodic in space and migrate at a constant velocity through the domain  \parencite{Kopell1973}. Such solutions have been observed in a wide range of PDE models, including but not limited to dynamics of dryland vegetation patterns \parencite{Sherratt2005}, intertidal mussel beds \parencite{Bennett2018}, hydrothermal waves \parencite{Hecke2003}, solar cycles \parencite{Proctor2000}, and pulses in excitable systems \parencite{Bordyugov2010}. We note that PTW also occur in integrodifferential equations \parencite{Gourley2001,Eigentler2023spike}, integrodifference equations \parencite{Kot1992,Britton1990}, and individual based models \parencite{Sherratt1996,Degond2022}. In this paper, our sole focus lies on PDE models on a one-dimensional space domain.

A remarkable feature of PDE models admitting PTWs is that they typically exhibit multistability of PTWs \parencite{Busse1978,Bastiaansen2018,Sherratt2021}. This means that, assuming a sufficiently large or infinite spatial domain, if one PTW is stable for a given set of PDE parameters, then other PTWs with different emergent properties (e.g., wavelength, wavespeed, wavenumber) are also stable. PTW stability in one-dimensional space domains is a well-explored topic: the stability of a PTW can be determined through a calculation of its essential spectrum. Results are typically visualised through the Busse balloon \parencite{Busse1978} which indicates regions of PTW stability in a two-dimensional parameter plane, spanned by the main PDE bifurcation parameter and one of the emergent properties of the PTWs (see \cref{fig: klausmeier busse spectra wavelength changes}A for an example).

Spatio-temporal patterns described by PTWs often occur in systems that undergo exogeneous change (e.g. climate change impact on dryland vegetation patterns) \parencite{UNGlobalLandOutlook2017}. It is thus crucial to completely understand how PTWs, and their stability, evolve under changing PDE parameters. Under changing parameters, PTWs preserve their wavelength as long as the PTW of that wavelength remains stable. Wavelength changes can only occur after a PTW crosses a stability boundary in the Busse balloon. This has important consequences, because wavelength changes typically cannot be reversed by simply reversing the parameter change. This is a well known feature, known as hysteresis \parencite{Sherratt2013}. The exact dynamics of how crossing a stability boundary leads to destabilisation remain underexplored. Nevertheless, it is known that the type of stability boundary, which is classified by the shape of the essential spectrum at the boundary (see e.g., \textcite{Sherratt2013a,Stelt2013} for a detailled overview) a PTW crosses upon destabilisation affects the dynamics: Eckhaus boundaries lead to a wavelength change, while Hopf boundaries can lead to oscillations of pattern peaks with the wavelength being preserved in the vicinity of the stability boundary \parencite{Dagbovie2014,Bennett2019}. 

However, even for Eckhaus stability boundaries, numerical results highlight that destabilisation is not instant upon crossing the stability boundary, and that PTWs can persist for biologically significant times after losing stability, exhibiting a \textit{delayed loss of stability}, before they eventually undergo a wavelength change \parencite{Sherratt2013,Sherratt2016b}. This highlights that stability boundaries in the Busse balloon do not provide sufficient information to predict wavelength changes of PTWs. Current evidence of this delayed loss of stability that induces a time delay between the crossing of a stability boundary in the Busse balloon and the occurrence of a wavelength change is anecdotal and descriptive, rather than predictive, and based on piecewise constant bifurcation parameter regimes only \parencite{Sherratt2013,Sherratt2016b}. It is worth noting, however, that there exists theory on delayed loss of stability phenomena in ODE systems \parencite{de2008smoothness,de2016entry,liu2000exchange,neishtadt1987persistence,kaklamanos2022entry} and theory which links the rate of change of a parameter to transient behaviour after crossing a bifurcation in non-PTW-admitting PDE systems \parencite{Dalwadi2023}.

In this paper, we focus on a novel predictive approach to quantify this delayed loss of stability. We develop a predictive understanding of the order of magnitude of the time delay that occurs between a PTW destabilisation at an Eckhaus stability boundary and the occurrence of an irreversible wavelength change. We show that the precise dynamics of parameter changes have a strong influence on when a wavelength change occurs and link these dynamics to the essential spectra of the PTWs. We develop this theory using a model describing dryland vegetation stripes. Further, we show that all results also apply to a model for intertidal mussel beds and thus argue that the predictive framework applies to all PTWs of PDE models that lose their stability at an Eckhaus stability boundary.

The paper is structured as follows. We describe the models used to obtain our results in \cref{sec: model}. In \cref{sec: Busse balloon insufficient}, we review the state-of-the-art knowledge on how essential spectra and the Busse balloon are used to predict wavelength changes of PTW, but also highlight why this information is not always sufficient. We use \cref{sec: time delay vs bif para} to provide information on how the delayed loss of stability between PTW destabilisation and wavelengh changes depends on the system's parameter values, but highlight that parameter values alone cannot predict wavelength changes. \cref{sec: time delay vs spectrum} contains the main result of this paper: the theory and practical implementation of a predictive method, based on the PTWs' essential spectra, that provides information on the order of magnitude of the the delayed loss of stability between PTW destabilisation and wavelength changes. We discuss the importance of our results in \cref{sec: discussion}.

\section{The model}\label{sec: model}

We consider the reaction-advection-diffusion system
\begin{align} \label{eq: general model}
	\frac{\partial u}{\partial t} &= f(u; \alpha) +  \nu \frac{\partial v}{\partial x} + d\frac{\partial ^2 u}{\partial x^2}, \quad x \in \R, t\ge 0.
\end{align}
Model densities are represented by $u  = (u_1, \dots, u_n) \in \R^n$. Non-spatial dynamics are accounted for by the function $f(u;\alpha) = (f_1(u;\alpha), \dots , f_n(u;\alpha))$, where $\alpha = (A, A_1, A_2, \dots) \in \R^n_+$ is a set of model parameters, and $A$ denotes the main bifurcation parameter of the system. Spatial dynamics comprise diffusion of all model densities with diffusion coefficient $d = (d_1, \dots d_n) \in \R^n_+$, and advection (at speed $\nu>0$) of one model density only, say density $u_n$, i.e., $\nu = (0, \dots, 0, \nu)$.

We assume the following on the function $f$ and the parameters $\nu$ and $d$:
\begin{itemize}
	\item Let $\overline{u}(\alpha)$ be a positive, spatially uniform equilibrium of \eqref{eq: general model}, i.e., $f(\overline{u}; \alpha) = 0$.
	\item Let $\overline{u}(\alpha)$ be stable for $A>A_H \in \R$ and lose its stability at a Hopf bifurcation at $A=A_H$.
	\item Let \eqref{eq: general model} admit stable PTW solutions for $A \in A_{\PTW} = [A_L,  A_U]$, where $A_U\ge A_H$.

A Periodic travelling wave (PTW) solution of \eqref{eq: general model} is a solution $U(z) = u(x,t), z=x-ct, c\in \R$ which satisfies 
\begin{align}
    0 = f(U;\alpha) + (c+\nu)\frac{\dif U}{\dif z} + d \frac{\dif^2 U}{\dif z^2}.
\end{align}
This travelling wave system is obtained from \eqref{eq: general model} through transformation into travelling wave coordinates, $z=x-ct$.

\item Let stability boundaries in the system's Busse balloon be of Eckhaus type.
\end{itemize}
In essence, the last two bullet points ensure the occurrence of PTWs that lose their stability at an Eckhaus boundary; this is the setting for which we aim to explore the concept of a delayed loss of stability in more detail. The first two bullet points provide us with a way to construct PTWs through numerical continuation (see \textcite{Sherratt2012,Sherratt2013a,Eigentler2020savanna_coexistence} for a detailed explanation).

Below, we present a mathematical model of pattern formation in dryland plant ecosystems that fits into model class \eqref{eq: general model} and satisfies the aforementioned hypotheses. This system will be used to visualise our results throughout the main text. Moreover, in \cref{sec: mussel model def}, we present a second model of class \eqref{eq: general model} for which we repeat our analysis to provide evidence that results presented in this paper apply to all models fitting into class \eqref{eq: general model}.

\subsection{The extended Klausmeier model for dryland vegetation patterns}

Throughout the main text of the paper, we use the \textit{extended Klausmeier model} to develop and illustrate our results. The model was first proposed by \textcite{Klausmeier1999} to describe vegetation stripes that form parallel to contours on gentle slopes in dryland ecosystems. For a detailed overview of the underlying ecological dynamics and other modelling approaches, we refer to the comprehensive reviews by \textcite{Gandhi2019a,Meron2016}. The nondimenionalised (see \textcite{Klausmeier1999,Sherratt2005} for the nondimensionalisation\footnote{The nondimensionalisations in these papers do not include $D=d_1/d_2$, which describes the ratio between the water diffusion coefficient $d_2>0$, and the plant diffusion coefficient $d_1>0$.}) model we consider in this paper is

\begin{subequations}
		\label{eq: klausmeier model}
		\begin{align}
            \frac{\partial u }{\partial t} &= \overbrace{u^2w}^{\text{plant growth}} - \overbrace{B u}^{\text{plant loss}} + \overbrace{\frac{\partial^2 u}{\partial x^2}}^{\text{plant dispersal}}, \\
			\frac{\partial w }{\partial t} &= \underbrace{A}_{\text{rainfall}} - \underbrace{w}_{\text{evaporation}} - \underbrace{u^2w}_{\substack{\text{water uptake} \\ \text{by plants}}} + \underbrace{\nu\frac{\partial w}{\partial x}}_{\substack{\text{water flow}\\ \text{downhill}}} + \underbrace{D\frac{\partial^2 w}{\partial x^2}}_{\text{water diffusion}}.	
		\end{align}
	\end{subequations} 
 The densities $u(x,t)$ and $w(x,t)$ describe the plant density and water density, respectively, at space point $x\in \R$ and time $t\ge0$. Water diffusion was not part of the system originally, but is a widely used addition (e.g. \parencite{Siteur2014,Zelnik2013,Eigentler2020_multispecies_pattern_math}) which leads the model to be referred to as the \textit{extended} Klausmeier model. Typically, the main bifurcation parameter of the model is the rainfall constant $A\ge0$, sincee it represents the environmental stress acting on the system. This model fits into the general framework \eqref{eq: general model} by setting (with a slight abuse of notation) $u=(u,w), \alpha = (A,B), d=(1,D)$, and $f(u;\alpha) = (u^2w-Bu, A-w-u^2w)$. The analysis of PTW solutions of the extended Klausmeier model representing vegetation stripes has a rich history (e.g., (\textcite{Bastiaansen2018, Bennett2019, Consolo2019, Consolo2019a, Marasco2014, Sherratt2010, Sherratt2005, Sherratt2011, Sherratt2013, Sherratt2013III, Sherratt2013IV, Sherratt2013V, Sherratt2007, Siero2018, Siero2019, Siteur2014,Wang2019, Wang2018a})) and thus the model is an ideal example to investigate the delayed loss of stability property. Unless otherwise stated we use $B=0.45, \nu = 182.5, D=500$ \parencite{Klausmeier1999,Siteur2014}.

\section{The Busse balloon provides insufficient predictions of wavelength changes}\label{sec: Busse balloon insufficient}

\begin{figure}
	\includegraphics[]{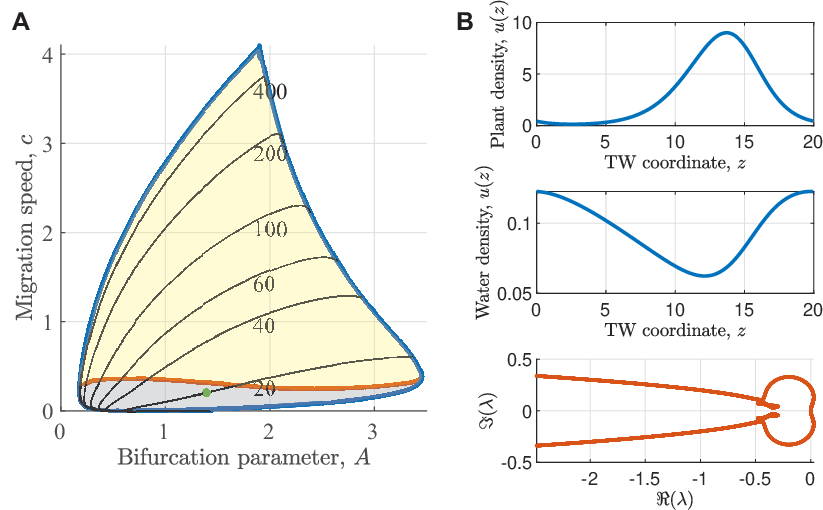}
	\caption{\textbf{Busse balloon and essential spectra.} \textbf{A:} Busse balloon of the Klausmeier model \eqref{eq: klausmeier model}. Shaded regions visualise regions of pattern existence, split into stable (yellow) and unstable (grey) patterns. Existence boundaries are shown in blue, stability boundaries in red. Annotated solid black curves show wavelength contours. The green dot on the $L=20$ contour indicates the location of the solution shown in \textbf{B}. \textbf{B:} One period of an example PTW for $A=1.4$ and $L=20$ (top and centre). Its essential spectrum is shown in the bottom panel; notice that the essential spectrum trespasses in the region $\Re(\lambda)>0$. Other parameter values are $B=0.45, \nu = 182.5, D=500$ across both figures.}\label{fig: klausmeier busse spectra wavelength changes}
\end{figure}

It is a well known feature of models of class \eqref{eq: general model} that multistability of PTW solutions occurs \parencite{Busse1978,Bastiaansen2018,Sherratt2021}. That is, if for a set of PDE parameters $\alpha = \alpha^\ast$, \eqref{eq: general model} admits a stable PTW solution with the wavelength-wavespeed-wavenumber triple $(L,c,k)=(L^\ast,c^\ast,k^\ast)$, then, for the same PDE parameters  $\alpha = \alpha^\ast$, other PTW solutions (forming a continuum if the domain is infinite) with different emergent properties ($L\neq L^\ast$, $c\neq c^\ast$, $k\neq k^\ast$) are also stable\footnote{The exception is the location of a Turing-Hopf bifurcation, at which, if supercritical, there is only one stable PTW.}. This information can be neatly summarised visually through a stability diagram, often termed the \textit{Busse balloon} \parencite{Busse1978}, see e.g. \Cref{fig: klausmeier busse spectra wavelength changes}A, in a parameter plane spanned by the main bifurcation parameter of the PDE system ($A$), and one of the PTW's emergent properties (here $c$).

To construct a Busse balloon, information on PTW stability is required. PTW stability is determined by its essential spectrum (\Cref{fig: klausmeier busse spectra wavelength changes}B). Practically (for full details see \textcite{Rademacher2007,Sherratt2012,Sherratt2013a}), the essential spectrum $\Lambda \subset \C$ of a PTW describes the growth rate (to linear order) of perturbations to the PTW. Thus, if $\mu:=\max_{\lambda\in\Lambda, \lambda \neq 0}(\Re(\lambda))<0$, the corresponding PTW is stable, and unstable otherwise. The origin $\lambda=0 \in \Lambda$ is excluded from this stability definition, as it is always part of the essential spectrum due to the translation invariance of PTWs.

\Cref{fig: klausmeier busse spectra wavelength changes}A shows the Busse balloon for the Klausmeier model for a specific choice of the model parameters (see the caption for the precise values). In particular, it visualises the Eckhaus stability boundary (red) which splits the PTW existence region (bounded by blue curves) into stable and unstable PTWs.  Thus, for any given wavelength, the intersection of the wavelength contour (black) with the stability boundary determines the value of the bifurcation parameter $A=\Astab$ at which the PTW loses its stability. Here, the stability boundary is of Eckhaus type (meaning, the essential spectrum evolves towards instability through a change in curvature at the origin; see \textcite{Sherratt2013a,Bennett2019,Dagbovie2014} for more information on types of stability boundaries). Intuitively, one would expect the PTW to undergo a wavelength change immediately upon crossing the boundary \parencite{Dagbovie2014}. However, in previous papers, a delayed loss of stability phenomenon has been highlighted. More precisely, it has been shown that, provided that the bifurcation parameter is varied at a sufficiently large rate in a step-wise manner, wavelength changes only occur well beyond the stability boundary, and with a time delay \parencite{Sherratt2013,Sherratt2016b}. 

\begin{figure}
	\includegraphics[width=\linewidth]{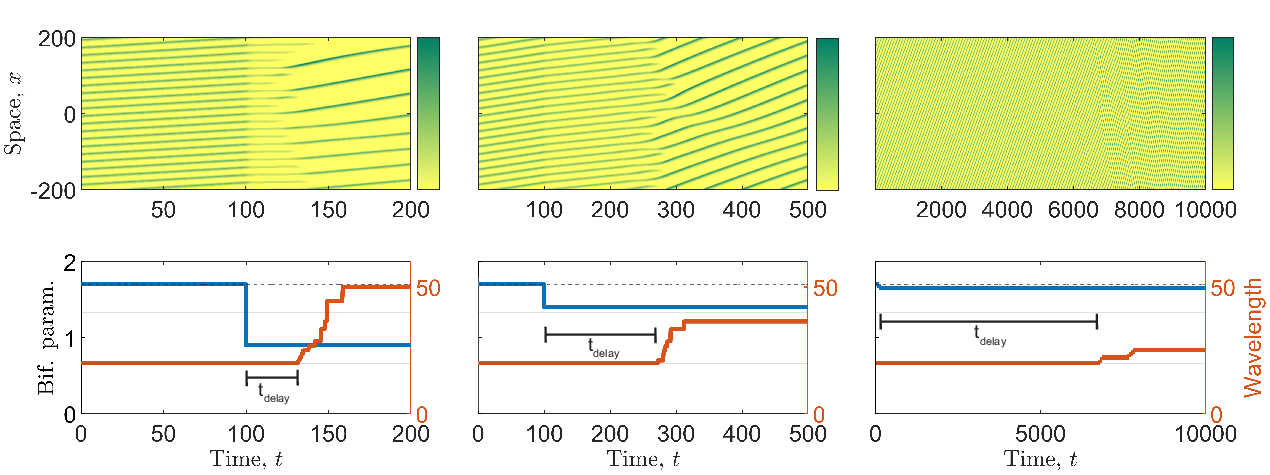}
	\caption{\textbf{Delay examples.} The top panel in each of the rows shows the contour plot of the plant density $u$ of system \eqref{eq: klausmeier model} in the time-space parameter plane. The simulation is initialised with a stable PTW constructed using numerical continuation. The bifurcation parameter (blue curve in bottom panel) is kept at its initial value for 100 time units before it is abruptly decreased to $A=\Atarget$ beyond the stability boundary. Here, $\Atarget = 0.9$ (left), $\Atarget = 1.4$ (middle), and $\Atarget = 1.65$ (right). A wavelength change (red curve in bottom panel) only occurs after a time delay $\tdelay$. Note the different limits on the time axes. In all three cases, $A_0=1.7$ and $\Astab\approx 1.69$. Other parameter values are $B=0.45, \nu = 182.5, D=500$ across all figures.}\label{fig: klausmeier delay examples}
\end{figure}

There is a possibility that these observations are the results of numerical errors that occur close to the stability boundary. However, we were able to independently verify this phenomenon using our numerical methods (\Cref{fig: klausmeier delay examples}). For this, we initialised simulations with a stable pattern (constructed using numerical continuation) located close to the stability boundary at $A=A_0$, where 
$0 < A_0-\Astab \ll 1$. Here, we chose a wavelength $L=20$ PTW at $A=1.7$ with the stability boundary being at $\Astab \approx 1.69$. After an initial calibration phase of 100 time units, we instantaneously changed the bifurcation parameter to $A=\Atarget < \Astab$, continued the simulation and recorded the time delay $\tdelay$ between crossing a stability boundary and the occurrence of a wavelength change. Significantly, this independent verification also revealed that the length of the delay can differ by several order of magnitudes depending on the value of $\Atarget$ (\Cref{fig: klausmeier delay examples}). The remainder of the paper aims to characterise why such an order of magnitude difference in the time delay $\tdelay$ exists and how the order of magnitude of such a delay can be predicted. 

Before proceeding to characterise the delayed loss of stability phenomenon in more detail, we remark that the delay is \textit{approximately memoryless} (relative to the order of magnitude difference reported for parameter changes) with respect to the dynamics that occur before crossing the stability boundary in the Busse balloon (\cref{sec: memoryless} and \Cref{fig: suppl: klausmeier memoryless}). Combined, this provides compelling - yet purely \textit{descriptive} - numerical evidence of a delayed loss of stability phenomenon, with the observed delays spanning several orders of magnitude depending on parameter values. This means that, under some parameter regimes, PTWs which are unstable according to the Busse balloon can be realised as transients over ecologically relevant timescales (\Cref{fig: klausmeier delay examples}). Therefore, the Busse balloon does not provide sufficient information on when wavelength changes occur in these cases. Below, we investigate this phenomenon further and develop more \textit{predictive} information on what determines the order of magnitude of wavelength change delays.

\section{Wavelength changes and their dependence on parameter distance to stability boundaries}\label{sec: time delay vs bif para}

\begin{figure}
	\includegraphics[]{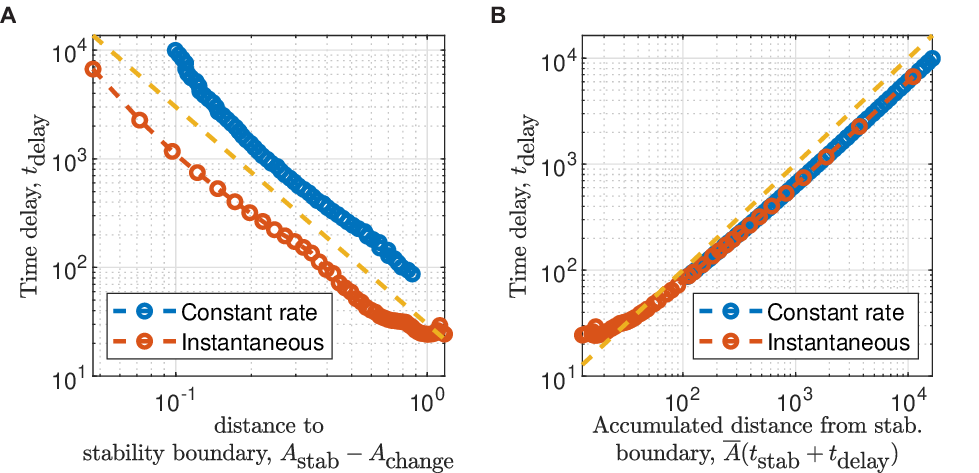}
	\caption{\textbf{Time delay in relation to bifurcation parameter changes.} \textbf{A}: The relation between the time delay $\tdelay$ and the distance of the bifurcation parameter $A$ from the stability boundary at the time of the wavelength change is shown for an instantaneous parameter change to a target value $\Atarget$ (red), and for a regime in which $A$ decreases at constant rate $m$ (blue) with each datapoint corresponding to a differnt value of $m$. The dashed lines have slope $-2$. \textbf{B}: The relation between the time delay $\tdelay$ and the accumulated distance from the stability boundary at the time of the wavelength change, $\overline{A}(\tstab + \tdelay)$, is shown for both parameter change regimes.  The dashed line has slope $1$. Other parameter values are $B=0.45, \nu = 182.5, D=500$.}\label{fig: delay vs A}
\end{figure}

Having established that wavelength changes do not occur instantaneously after crossing an Eckhaus stability boundary, we first quantified how the order of magnitude of the time delay $\tdelay$ depends on the distance of the bifurcation parameter to the stability boundary. To do so, we repeated our numerical simulations described in the previous section for a wide range of $\Atarget$.
To allow comparison to other parameter change regimes (see below), we denote the bifurcation parameter at which the wavelength change occurs by $\Ajump$. Here, $\Ajump = \Atarget$ due to the choice of the change regime of the bifurcation parameter. Our simulations revealed that changes of the bifurcation parameter to values further from the stability boundary (i.e. lower values of $\Atarget$) decrease the time delay between the parameter change and the wavelength change (\cref{fig: delay vs A}A, red). Moreover, we recorded that the time delay $\tdelay$ approximately scales with the distance to the stability boundary $\Astab-\Atarget$ through $\tdelay \sim (\Astab-\Atarget)^{-2} = (\Astab-\Ajump)^{-2}$. We thus conclude that the order of magnitude of the time delay $\tdelay$ is determined by the distance of the bifurcation parameter to the stability boundary.

The aforementioned numerical investigation, albeit useful for our understanding of the mechanisms underlying the delayed loss of stability, is rather unnatural from a biological point of view. In real-world ecosystems, changes of environmental conditions are rarely instantaneous. Rather, changes are often gradual. We therefore repeated our simulations with a regime in which the bifurcation parameter decreased linearly from its initial value $A_0 = 1.7$ after the initial calibration phase, i.e. $A(t)=A_0$ for $t\le 100$ and $A(t)=A_0 - mt$, for some $m>0$, for $t>100$. Note that, in contrast to the previous simulations, the value of the bifurcation parameter at which the wavelength change occurs ($\Ajump$) is an emergent property of the simulation rather than an input. Our simulations with this parameter regime revealed a strikingly similar relation between the time delay $\tdelay$ and the distance to the stability boundary at the wavelength change ($\Astab - \Ajump$) compared with the regime of instantaneous changes in the bifurcation parameter. Again, we observed that  $\tdelay \sim (\Astab-\Ajump)^{-2}$ (\cref{fig: delay vs A}B, blue). However, for fixed distance to the stability boundary, the time delays in the instantaneous change regime were much shorter (up to one order of magnitude) than in the constant rate of change regime (\cref{fig: delay vs A}).

The comparison of the two regimes above highlights that there is a clear qualitative relation between how far a PTW can cross a stability boundary and the order of magnitude of the time delay before a wavelength change occurs. However, the quantitative differences between the two parameter change regimes highlight that the distance to the stability boundary alone has little predictive power. Instead, we hypothesised that the distance to the stability boundary during the entirety of the delay phase must be accounted for. To test this hypothesis, we defined the \textit{accumulated distance from the stability boundary} as
\begin{align}\label{eq: aumulated distance}
	\overline{A}(t) := \int_{\tstab}^{t} A(\tau) \dif \tau, \quad t\ge \tstab
\end{align}
where $\tstab$ denotes the time at which the bifurcation parameter last crossed the stability boundary to push the PTW into an unstable regime. If our hypothesis regarding the predictive power of the accumulated distance from the stability boundary $\overline{A}(t)$ was true, then we would see no dependence of the time delay $\tdelay$ on $\overline{A}(\tstab+\tdelay)$. We compared the accumulated distance from the stability boundary at the time of the wavelength change, $\overline{A}(\tstab + \tdelay)$ with the time delay $\tdelay$ for both previously described parameter change regimes. Most significantly, we discovered that there is a clear relation between the accumulated distance from the stability boundary and the time delay and thus rejected our hypothesis on the predictive power of accumulated distance from the stability boundary. However, we also found excellent quantitative agreement across the data from both parameter change regimes (\cref{fig: delay vs A}C). Moreover, we detected that the relation between the time delay and the accumulated distance from the stability boundary  is approximately $\tdelay \sim \overline{A}(\tstab + \tdelay)$. These numerical data show that the order of magnitude of the time delay $\tdelay$ is determined by the accumulated distance from the stability boundary $\overline{A}(\tstab + \tdelay)$, independent of the parameter change regime. Yet, the implicit nature of the relationship does not provide any predictive information on the order of magnitude of the time delay of any wavelength change.

\section{The maximum real part of the essential spectrum determines the order of magnitude of the delay}\label{sec: time delay vs spectrum} 

\begin{figure}

\begin{tikzpicture}
\node at (0,0) {\includegraphics[width=\linewidth]{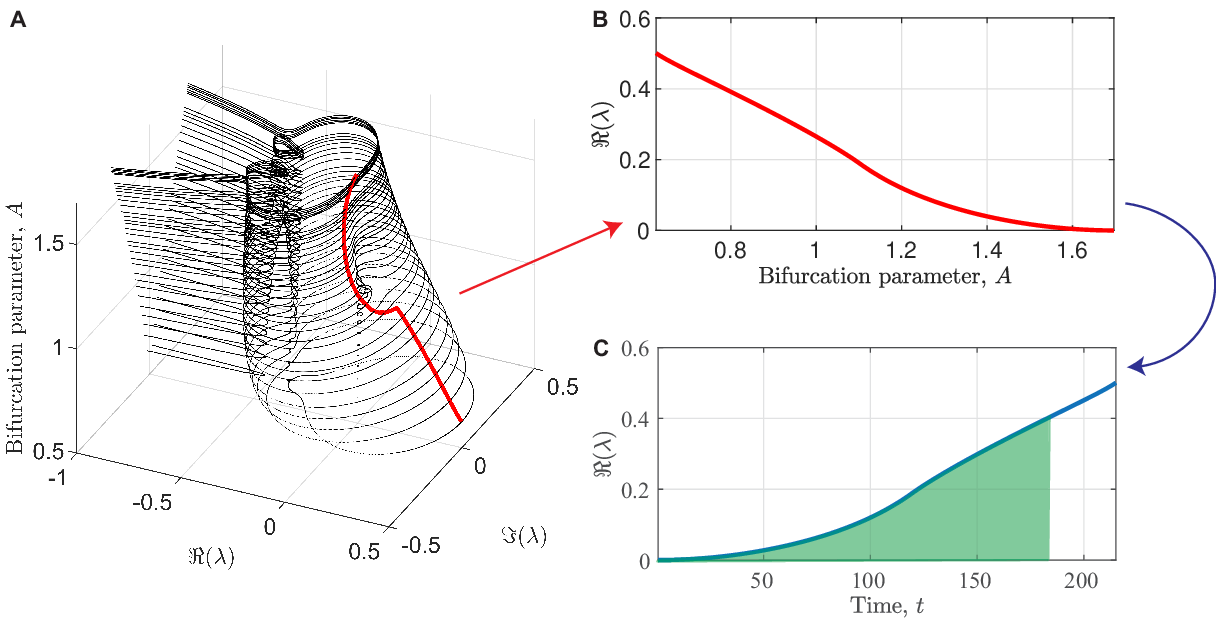}};
\node at (6.25,-0.1) {$A=A(t)=A_0-mt$};
\node at (5,-2.75) {$\overline{\mu}(t^*)$};
\node at (6,-3.6) {$t^*$};
\end{tikzpicture}

	\caption{\textbf{Accumulated maximal instability.} \textbf{A}: Spectra of PTW with wavelength $L=20$ of system \eqref{eq: klausmeier model} are shown as a stack with the bifurcation parameter $A$ on the z-axis. The red curve traces the maximum real parts of the spectra. Note that if the maximum occurs away from the real axis, the maximum is not unique due to the occurrence of complex conjugates. Only the maximum occurring in the $\Im(\lambda) <0$ plane is shown. \textbf{B}: The maximum real part of the spectra (red curve in \textbf{A}) is plotted against the bifurcation parameter $A$. \textbf{C}: The maximum real part of the spectra is plotted against time. The curve is obtained by transforming the x-axis in \textbf{B} using the relation $A=A(t)$. Here, $A(t) = A_0 - mt$ with $A_0 = 1.7$, $m=0.005$. The green shaded area indicates how the accumulated maximum instability $\mubar(t^\ast))$ is calculated for a specific value $t=t^\ast$. Other parameter values are $B=0.45, \nu = 182.5, D=500$.}\label{fig: accumulated max instability}
\end{figure}
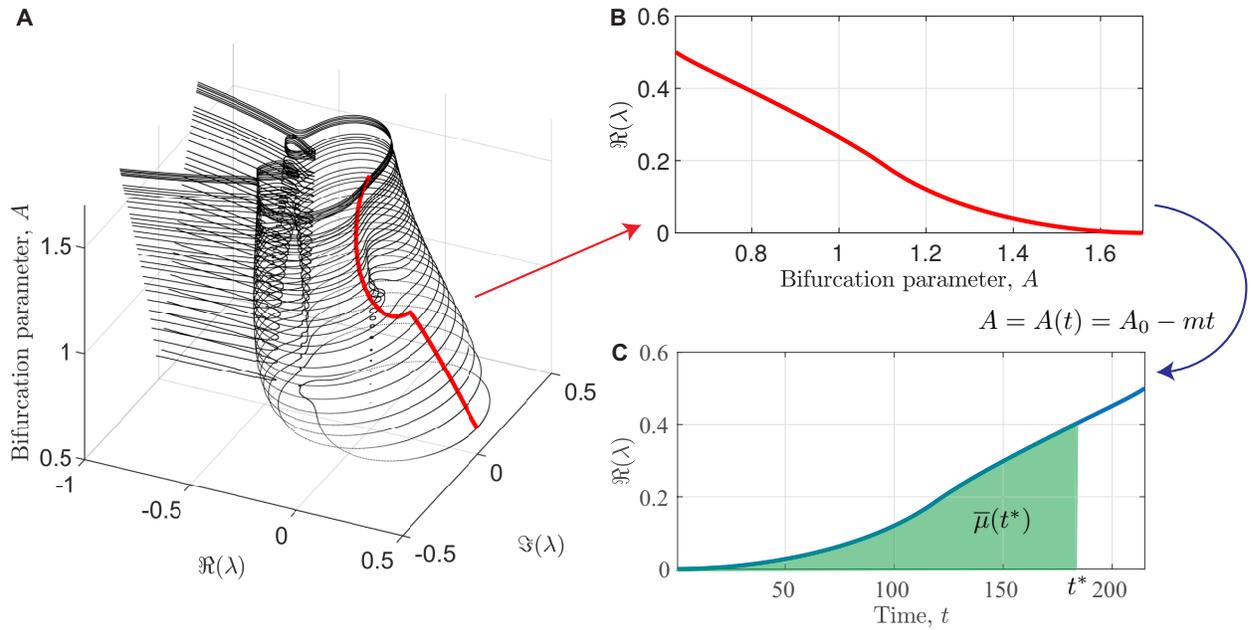

\begin{figure}
	\includegraphics[]{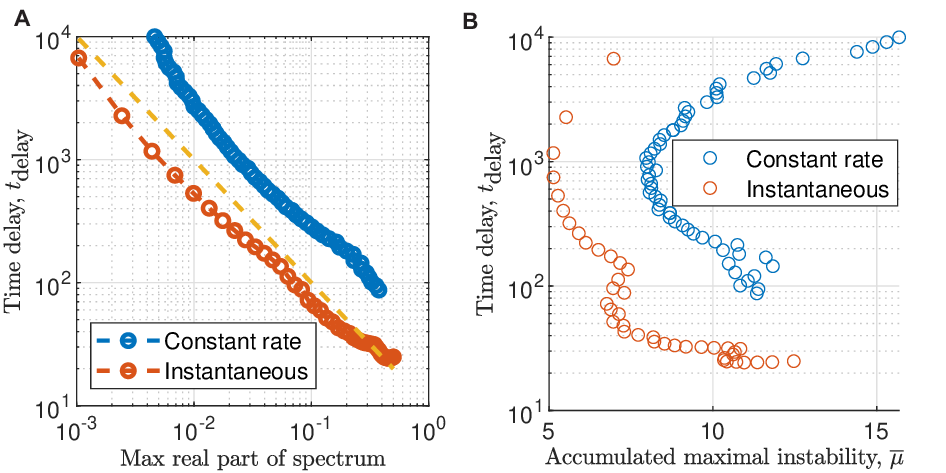}
	\caption{\textbf{Time delay of wavelength change in relation to the maximum real part of the essential spectrum.} \textbf{A}: The time delay of a wavelength change that occurs after crossing a stability boundary is compared with the maximum real part of the essential spectrum of the unstable solution at which the wavelength change occurs for instantaneous changes of the bifurcation parameter (red), and constant rates of change of the bifurcation parameter (blue). The dashed line has slope -1. \textbf{B}: The time delay is compared with the accumulated maximal instability for both parameter change regimes. Other parameter values are $B=0.45, \nu = 182.5, D=500$.}
	\label{fig: delay vs max real part spectrum}
\end{figure}

The previous section revealed a clear relation between the time delay $\tdelay$ before a wavelength change occurs after a PTW crosses a stability boundary and the bifurcation parameter's accumulated distance from the stability boundary. However, this relation possesses no predictive power because both these quantities depend on the time delay. Nevertheless, the relation highlights the importance of considering the details of solution dynamics during the entirety of the delay phase, rather than solely focussing on the wavelength change itself. 

Given that the model's parameter values alone hold no predictive power on wavelength changes, we next turned our attention to the essential spectra of the destabilised PTWs. As described above, the essential spectrum of a PTW determines the behaviour (up to linear order) of perturbations to the PTW, and therefore provides information on the linear stability of the PTW. An example spectrum is shown in \cref{fig: klausmeier busse spectra wavelength changes}B and a stack of spectra for different values of $A$ that were obtained for PTWs with wavelength $L=20$ and is shown in \cref{fig: accumulated max instability}A. The maximum real part of the essential spectrum, $\mu(A) := \max_{\lambda\in\Lambda, \lambda \neq 0}(\Re(\lambda(A))$ (red curve in \cref{fig: accumulated max instability}A) is of particular significance because it describes the linear growth rate of the fastest growing perturbation of the form $\overline{U} \sim \exp(\lambda t)$ to the underlying PTW. We therefore hypothesised that given the previously reported relation between time delay $\tdelay$ and bifurcation parameter $A$, the delay and the spectra should be related through $\tdelay \sim \mu(\Ajump)^{-1}$. This is in part motivated by general theory (e.g. \textcite{Hastings2018}) and dryland vegetation pattern-specific results \parencite{Eigentler2019Multispecies} on transients which highlight that unstable states whose leading eigenvalues feature a small but positive real part can lead to the preservation of intrinsically unstable states over ecologically relevant timescales.

To test this hypothesis, we returned to the numerical data for the two different parameter change regimes investigated in the previous section and calculated the essential spectra for the PTWs at the parameter values at which they underwent a wavelength change. To do so, we implemented the numerical continuation method developed by \textcite{Rademacher2007} (but see also \textcite{Sherratt2012,Sherratt2013a}) in AUTO-07p \parencite{AUTO}. For both parameter change regimes, we found that indeed $\tdelay \sim \mu(\Ajump)^{-1}$ (\cref{fig: delay vs max real part spectrum}A). Thus, the order of magnitude of the time delay $\tdelay$ is affected by how far the spectrum of the unstable PTW extends beyond the imaginary axis in the complex plane.

Despite the clear relation between the time delay $\tdelay$ and the maximum real part of the essential spectrum of the unstable PTW at the wavelength change, $\mu(\Ajump)$ in both parameter change regimes, there were significant (up to one order of magnitude) differences across the two regimes. This highlighted that the spectrum at the wavelength change alone does not possess sufficient power to predict the order of magnitude of the time delay. Motivated by the results of the previous section, we instead considered the maximum real parts of the essential spectra of the PTWs during the entirety of the delay phase. Similar to the previous section, we defined the \textit{accumulated maximal instability} 
as
\begin{align}\label{eq: accumulated maximal instability}
	\mubar(A(t)) = \int_{\tstab}^t \mu(\tau)d\tau, \quad t\ge \tstab.
\end{align}

As in definition \eqref{eq: aumulated distance}, we highlight that $\tstab$ denotes the \textit{last}  time the system crossed a stability boundary from a stable to an unstable regime. We compared the accumulated maximal instability at a wavelength change, i.e. $\mujump:=\mubar(\Ajump)$, with the time delay $\tdelay$ for both parameter change regimes. In both cases, we found that there exists no clear relation between the accumulated maximal instability and the time delay (\cref{fig: delay vs max real part spectrum}C). Moreover, we observed that wavelength changes occur within a small interval $I_{\mujump}$, i.e. when $\mujump \in I_{\mujump}$. In our data, we observed wavelength changes within $I_{\mujump} \approx [5,16]$. While this does not provide us with one critical value of $\mujump$ at which the wavelength change occurs, the interval's range (in terms of order of magnitude) is much smaller than the range of the maximum real parts of spectra recorded at the wavelength changes (these spectra range from $10^{-3}$ to $5\cdot 10^{-1}$ in our data). We thus conclude that we can obtain an order of magnitude prediction of how long after crossing a stability boundary a PTW undergoes a wavelength change by tracking the order of magnitude of the accumulated maximal instability $\mubar$ until $\mubar \in I_{\mujump}$. We detail the procedure in the next section.

\subsection{Delay predictions in practice}

The previous section revealed that wavelength changes of PTWs after crossing a stability boundary occur when the accumulated maximal instability $\mubar(t)$ reaches the interval $I_{\mujump}$. The predictive power of this result can be exploited as follows. Consider a stable PTW of interest and a bifurcation parameter change regime $A(t)$ that pushes the PTW across a stability boundary at $t=\tstab>0$. Then, $\mubar(A(t))$ can be calculated by computing the essential spectra of the PTW along the wavelength contour it follows under the parameter change regime $A(t)$. The quantity $\mubar(t)$ can be examined over time and a prediction of the time delay $\tdelay$ can be made by determining the order of magnitude of the time $t$ at which $\mubar(t) \in I_{\mubar}$. Recall that for our data, we observed $I_{\mujump} \approx [5,16]$. For our predictive framework, we therefore chose $\mubar(t) = 10$ as the critical threshold. Using any other value within $I_{\mujump}$ would not have changed the order of magnitude of the predicted delay $\tdelay$. 

\begin{figure}
	\includegraphics{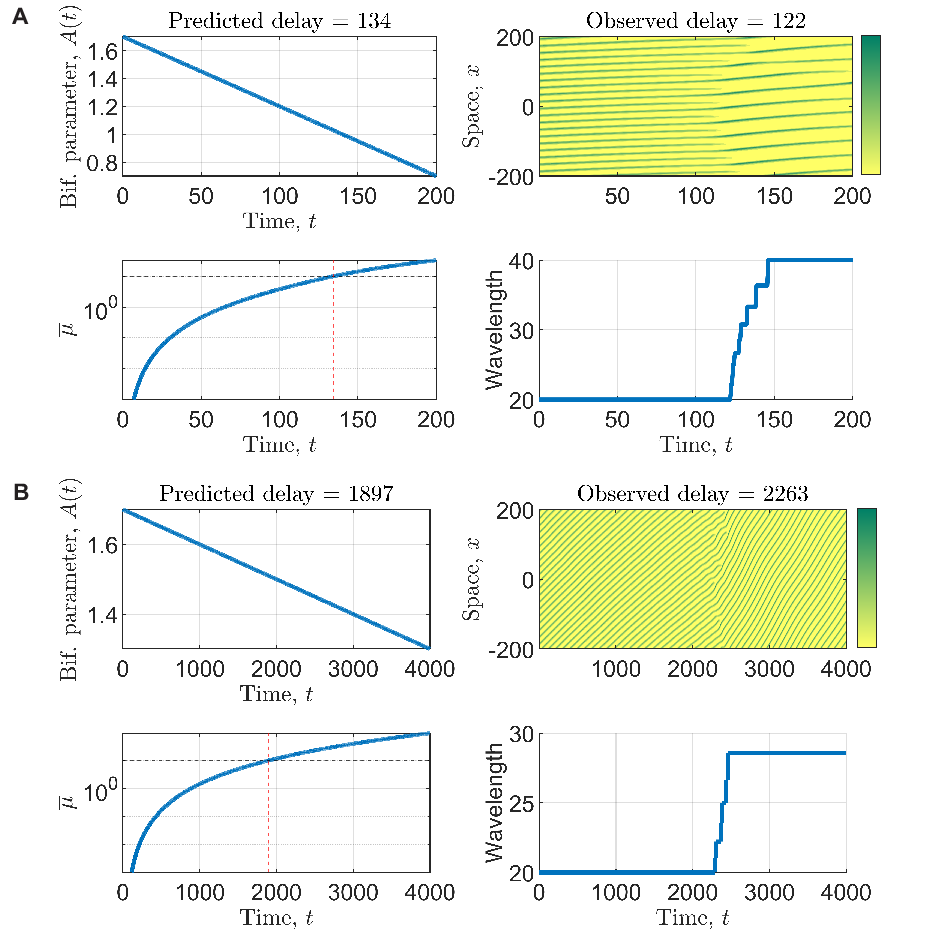}
	\caption{\textbf{Examples of delay predictions.} \textbf{A,B}: Two predictions of the time delay with comparisons to the observed delay are shown. In both parts, the top left panel shows the changing bifurcation parameter over time. The bottom left panel shows the accumulated maximal instability. The black dashed line shows the critical prediction threshold $\mubar=10$. The dashed red line shows the predicted time delay based on the $\mubar$ curve crossing the prediction threshold $\mubar=10$. The top right panel shows the simulation outcome in the time-space plane. The bottom right panel shows the observed wavelength in the simulation over time. The rates of change of the bifurcation parameter are $m=0.005$ in \textbf{A}, and $m=0.0001$ in \textbf{B}. Other parameter values are $B=0.45, \nu = 182.5, D=500$.}\label{fig: delay predictions examples}
\end{figure}

\cref{fig: delay predictions examples} shows two examples of the predictor with comparisons to numerical simulations. \cref{fig: delay predictions examples}A shows a prediction for a wavelength $L=20$ PTW in a regime in which the bifurcation parameter decays at a constant rate from an initial value close to the stability boundary, i.e. $A(t) = A_0-mt$ with $m = 0.005$. The prediction is compared with a numerical simulation, initialised at $A=A_0=1.7$ with a PTW constructed through numerical continuation. \cref{fig: delay predictions examples}B shows a prediction for the same PTW, but under a slower change regime $A(t) = A_0-mt$ with $m = 0.0001$. In both cases, there is excellent agreement between the order of magnitude of the delay prediction and the order of magnitude of the observed delay in a numerical simulation. We note that we tested the predictive method for other parameter change regimes and found similarly excellent agreement in all cases (\cref{fig: suppl: more delay predictions}).

The predictive framework is not only capable of predicting the time delay between a solution trajectory crossing a stability boundary in the Busse balloon and the occurrence of a wavelength change, but can also indicate to what extent solution changes in response to changes of the bifurcation parameter are reversible. Systems of type \eqref{eq: general model} that admit PTWs are known to feature hysteresis \parencite{Sherratt2013}. That is, wavelength changes that occur due to a decrease of the bifurcation parameter, cannot be reversed by simply reversing the changes to the bifurcation parameters. In the past, the stability boundaries in the Busse balloon have often been used to define critical thresholds that cause such irreversible changes \parencite{Sherratt2013a,Bastiaansen2020,Stelt2013}. However, the analysis in the preceding sections highlights that wavelength changes do not necessarily occur at the Busse balloon's stability boundary because a delay phase may occur. Our predictive framework is therefore able to characterise parameter change regimes that only cause reversible solution changes by characterising the accumulated maximal instability of these parameter changes.

\begin{figure}
	\includegraphics[width=\linewidth]{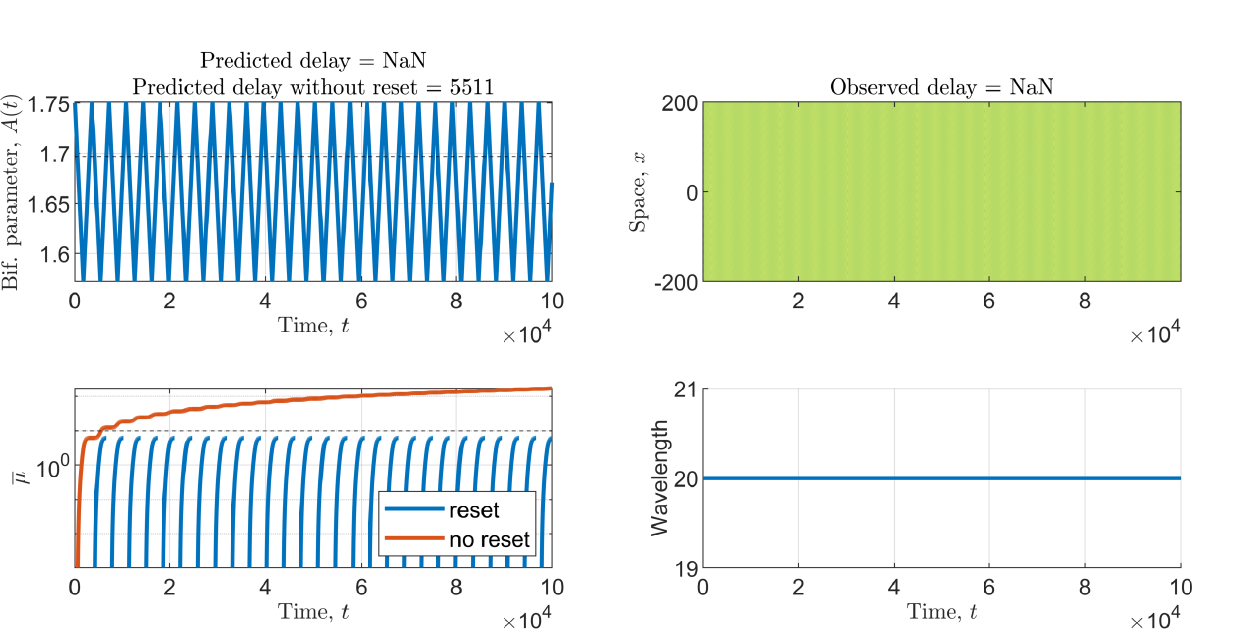}
	\caption{\textbf{Delay predictions reset in stable regions.} This simulation highlights that it is essential to reset the value of the accumulated maximal instability $\mubar$ whenever parameter changes push the system back into a stable region (blue in bottom left panel). If $\mubar$ is not reset (red in bottom left panel), then this leads to inaccurate predictions of the occurrence of wavelength changes. For a full description of the figure panels, see \cref{fig: delay predictions examples}; for a full description of the parameter change regime, see the main text. Other parameter values are $B=0.45, \nu = 182.5, D=500$.}\label{fig: delay prediction reset}
\end{figure}

For the characterisation of the reversibility of parameter change regimes, it is essential to recall that in the definition of the accumulated maximal instability $\mubar$ in \eqref{eq: accumulated maximal instability}, the quantity $\tstab$ refers to the \textit{last} time the bifurcation parameter crossed a stability boundary from a stable into a unstable regime. That is, the accumulated maximal instability is memoryless to previous delay phases and must be reset whenever parameters return into a stable regime. We highlight this property through the following counterexample, visualised in \cref{fig: delay prediction reset}, in which not resetting $\mubar$ leads to an inaccurate prediction of a wavelength change. For this, we initiated a model simulation with a wavelength $L=20$ PTW at $A=A_0=1.75$. We then varied the bifurcation parameter $A$ through the following periodic regime. First, we decreased $A$ at rate $m=10^{-4}$ for $t^\ast = 1800$ time units, i.e. $A(t) = A_0-m t$ for $2(n-1)t^\ast < t\le (2n-1)t^\ast, n\in \N$. We then increased $A$ at the same rate for the next $t^\ast$ time units, i.e. $A(t) = A(t^\ast) + m t$ for $(2n-1)t^\ast<t\le 2nt^\ast, n\in\N$. This periodic regime caused the system to spend time in both the stable and unstable regime for the $L=20$ PTW during each oscillation. We recorded (i) the accumulated maximal instability $\mubar$ as defined by \eqref{eq: accumulated maximal instability} where $\tstab$ denotes the \textit{last} time the parameters transitioned into an unstable regime and $\mubar$ is reset to $\mubar = 0$ whenever a transition to a stable regime occurs; and (ii) the accumulated maximal instability without reset, whose definition was identical to that of $\mubar$, with the exception that $\tstab$ denotes the \textit{first} time the parameter regime enters an unstable region.  If we assumed that the accumulated maximal instability $\mubar$ never resets, then our predictive framework would predict a wavelength change to occur within the first few oscillations due to an assumed additive effect of the maximal instability across the oscillations (\cref{fig: delay prediction reset} bottom left; red). However, as shown in the simulation of this parameter change regime in \cref{fig: delay prediction reset} (top right), and predicted by $\mubar$ with resets (\cref{fig: delay prediction reset} bottom left; blue), no wavelength change occurs. This highlights that the process is memoryless: the PTW ``forgets'' its excursion in the unstable regime as soon as it re-enters the stable part of the Busse balloon. We note that we have chosen a parameter regime in which the excursions into stable regions last for a sufficiently long time and are sufficiently far away from the stability boundary. We have not attempted to define the meaning of the term ``sufficiently'' in this context, but argue that an exploration of this is an important aspect of future work.

\section{Discussion}\label{sec: discussion}

Parameter changes in PDE systems admitting PTWs often cause a cascade of transitions between PTWs of different wavelengths \parencite{Rietkerk2021}. Wavelength changes are typically associated with PTW destabilisation when a PTW crosses a stability boundary in the Busse balloon \parencite{Bastiaansen2018}. However, previous work has noted a delayed loss of stability phenomenon \parencite{Sherratt2016b,Sherratt2013}. That is, there is a time delay between the crossing of a stability boundary and the occurrence of a wavelength change. Such delays have only been recorded for piecewise constant parameter change regime and all reports have been purely descriptive; they were noted as an aside when focussing on other research questions \parencite{Sherratt2013,Sherratt2016b}. In this paper, we have developed a predictive tool to determine the order of magnitude of the time delay between the crossing of a stability boundary and the occurrence of a wavelength change. 
Moreover, we present strong evidence that our predictive scheme applies to any parameter change regime and any PTW that loses its stability at an Eckhaus stability boundary.

A predictive understanding of wavelength changes affecting PTWs is of crucial importance. This is because PDE systems admitting PTWs exhibit hysteresis and wavelength changes of PTWs cannot be reversed by simply reversing the parameter change that has occurred \parencite{Sherratt2013,Stelt2013,Siero2019,Hardenberg2001,Meron2004}. Rather, a much larger change of the bifurcation parameter would be required to revert the system to its original wavelength. Thus wavelength transitions are examples of a tipping point in the sense that the PDE solution undergoes an almost instantaneous wavelength change after a long period of wavelength conservation (\cref{fig: klausmeier delay examples}). 

We note that there exists some confusion about the notion of tipping points in relation to PTWs. Pattern formation is sometimes attributed to be a mechanism to avoid tipping in the sense that it prevents the disappearance of a model density \parencite{Rietkerk2021,PintoRamos2023}. Here, following \textcite{Bastiaansen2018,Bennett2019}, we describe wavelength changes as a type of tipping. In the past, predictions of such tipping points for PTWs have solely relied on the location of stability boundaries in the Busse balloon \parencite{Sherratt2013a,Bastiaansen2020,Stelt2013}. However, the observation of a delayed loss of stability phenomenon which can last for significant timescales (\cref{sec: Busse balloon insufficient}) highlights that the Busse balloon provides insufficient information for predicting the occurrence of a wavelength change. Rather, we highlight that a sufficient ``amount'' of instability, quantified by the notion of the accumulated maximal instability that is based on maximum real part of the essential spectrum of a unstable PTW, needs to accumulate before a wavelength change occurs (\cref{sec: time delay vs spectrum}). Significantly, we show that this notion of the critical ``amount'' of the instability is identical across different parameter change regimes (\cref{fig: delay vs max real part spectrum}B) and even across different PDE systems (\cref{sec: mussel results,fig: suppl: mussels delay vs max real part spectrum}). From an ecological point of view, our results thus also add to the list of ``early warning signals'' that are used to detect tipping points, including wavelength transitions, before they occur \parencite{Scheffer2009,Kefi2014,Dakos2011}.

Despite our claim of the wide applicability of our results, it is important to emphasise that we developed our theory only for PTWs that lose their stability at an Eckhaus stability boundary. Models admitting PTWs can also feature Hopf stability boundaries (e.g., \parencite{Dagbovie2014,Bennett2019,Eigentler2020savanna_coexistence}; the type of stability boundary is determined by \textit{how} the essential spectrum crosses the imaginary axis (see \textcite{Sherratt2013a,Stelt2013} for more information). Previous studies have focussed on examining the different impacts of those two types of stability boundaries on PTW dynamics. While destabilisations of PTWs due to Eckhaus stability boundaries lead to wavelength changes (after a time delay), PTWs that are destabilised at a Hopf stability boundary preserve their wavelength and instead feature alternating oscillations in the pattern peaks provided parameters stay close to the stability boundary \parencite{Dagbovie2014,Bennett2019}. While Hopf-destabilised PTWs can also undergo a wavelength change \parencite{Eigentler2020savanna_coexistence}, we are not aware of any comprehensive understanding of what causes wavelength changes in these cases. We view such an understanding as an essential precursor before being able to characterise the (potential) occurrence of a delayed loss of stability phenomenon and have therefore focussed solely on PTW that lose their stability at Eckhaus boundaries.

Our analysis focussed on PDE models admitting PTW due to their wide applicability in ecology \parencite{Bennett2019,Sherratt2005}, fluid dynamics \parencite{Hecke2003}, magnetohydrodynamics \parencite{Proctor2000}, and excitable systems \parencite{Bordyugov2010}, among others. Nevertheless, delays in the loss of stability are not exclusive to PTWs, and it is therefore tempting to draw a parallelism to certain classes of ordinary differential equations (ODEs) \parencite{de2008smoothness,de2016entry,liu2000exchange,neishtadt1987persistence}. Recently, the role of the maximum eigenvalue in this context has been proven for a minimal system \parencite{kaklamanos2022entry}. Whereas previous theorems concerning the delayed loss of stability of singularly perturbed systems of ODEs assumed a complete separation of the eigenvalues (i.e., the eigenvalue which changes sign, and thus causes the loss of stability, should not intersect the other eigenvalues of the Jacobian), in \textcite{kaklamanos2022entry}, the authors prove that in cases in which eigenvalues do intersect, the maximum eigenvalue throughout the dynamics characterizes and determines the delay. This first result motivates further research in this direction. However, a major difference to the work presented in this paper is that systems of ODEs do not exhibit loss of memory like the PTW analysed in this paper: indeed, orbits which remain stable for longer periods are destabilized later, as the ``accumulated stability'' takes longer to be balanced by the ``accumulated instability'' \textcite{de2008smoothness,de2016entry}. Moreover, also for PDEs, the phenomenon of delayed loss of stability has been reported for stationary patterns in a two-dimensional model for dryland vegetation patterns, and characterised using Fourier analysis for a select number of wavelength changes \parencite{asch2023slow}. Combined, we argue that future work should reconcile these approaches with the aim of developing a general understanding of the delayed loss of stability across a wide range of types of continuum models.

Moreover, we highlight that PTWs also occur in integrodifferential equations \parencite{Gourley2001,Eigentler2023spike}, integrodifference equations \parencite{Kot1992,Britton1990}, and individual based models \parencite{Sherratt1996,Degond2022}. Theory for studying PTWs is best developed in PDE settings and, therefore, much less attention is currently being paid to other model types. In particular, we are not aware of tools to investigate PTW stability in any other model type. A characterisation of a delayed loss of stability phenomenon in these model types is a pressing, albeit challenging topic for future work. 

The fact that all our observation and results in this paper are strongly tied to the essential spectrum, which is a first order approximation of the stability of a given PTW \parencite{Rademacher2007}, provides a further important question for future work. Considering that we are observing the evolution of solutions in highly non-negligible intervals of time, higher order components might very well play a pivotal role. This leads to a considerably challenging question:
would the essential spectrum calculated to higher order approximations provide sufficient additional information on the delayed loss of stability of PTWs to justify the expected cumbersome increase in computations and complexity? A nonlinear approach could also increase the accuracy of our predictive method. Here, we presented a method that provides an order of magnitude prediction of the delay between PTW destabilisation and occurrence of a wavelength change. However, more quantitative estimates would provide deeper insights into wavelength changes of PTWs. 

Finally, we remark that this paper focusses only on the dynamics between pattern destabilisation and occurrence of a wavelength change. Crucially, it does not provide any information on the dynamics of the wavelength change itself. Characterising which new wavelength is chosen upon a PTW wavelength change is another underexplored question. Previous studies only highlighted significant differences between different model systems, ranging from small, almost gradual wavelength changes to period doubling regimes \parencite{Sherratt2016b}. Moreover, the various numerical simulations presented in this paper highlight that even within the same model, large differences between newly selected wavelengths occur depending on the parameter change regime. Moreover, our small number of simulations indicate that the time delay and the size of the wavelength change are inversely correlated (\cref{fig: delay predictions examples,fig: suppl: mussels busse spectra wavelength changes}). We thus hypothesise that, among other properties, the time delay that occurs due to a delayed loss of stability phenomenon plays are crucial role and we therefore view the results presented in this paper as an important precursor for this analysis.\\

\noindent \textbf{Acknowledgments:} 
L.E. was supported by the German Research Foundation (DFG) as part of the CRC TRR 212 (NC$^3$) – Project number 316099922. \\
M.S. was supported by the Italian Ministry for University and Research (MUR) through the PRIN 2020 project ``Integrated Mathematical Approaches to Socio-Epidemiological Dynamics'' (No. 2020JLWP23). \\

\noindent \textbf{Data availability:} Computational code used to obtain the results presented in this paper has been deposited in a Github repository which has been archived through Zenodo \parencite{delayedlossofstab_code_repo}.\\

\noindent \textbf{Author contributions:} \\
\textbf{L.E.:} Conceptualization, Software, Formal analysis, Investigation, Data Curation, Writing - Original Draft, Writing - Review \& Editing, Visualization, Project administration\\
\textbf{M.S.:} Conceptualization, Investigation, Formal analysis, Writing - Original Draft, Writing - Review \& Editing, Project administration

\printbibliography
\clearpage

\appendix

\section{Approximate memorylessness of the delay}\label{sec: memoryless}
\begin{figure}
	\centering
	\includegraphics[width=\linewidth]{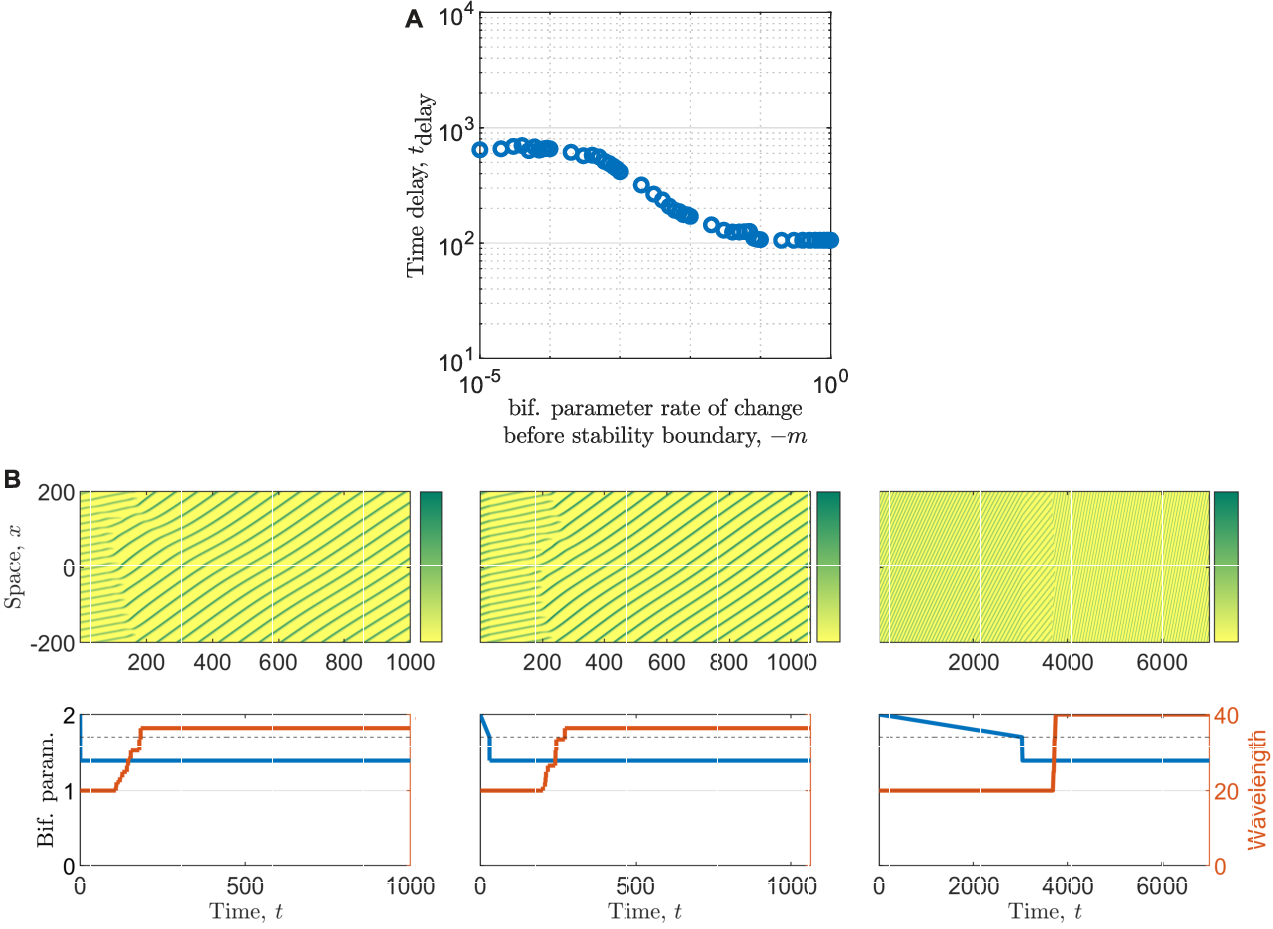}
	\caption{\textbf{Approximate memorylessness of the delay.} \textbf{A}: The time delays, $\tdelay$, for simulations with varying rate of change ($-m$) of the bifurcation parameter before hitting the stability boundary ($A=2 - mt$ for $t<\tstab$), and subsequent instantaneous switch to $A=\Atarget=1.4$. Note that the delay is approximately independent of $m$ (compared to the order of magnitude differences reported in \cref{fig: klausmeier delay examples}). \textbf{B}: The three panels show the simulation results for three examples of the data shown in \textbf{A} for $m=10^{-4}$ (left), $m=10^{-2}$ (middle), $m=1$ (right). Note the different limits on the time axes. Other parameter values are $B=0.45, \nu = 182.5, D=500$.}\label{fig: suppl: klausmeier memoryless}
\end{figure}

We obtained data on the approximate memorylessness of the delay $\tdelay$ with respect to the system dynamics before hitting a stability boundary as follows. We initiated each model simulation of system \eqref{eq: klausmeier model} with a PTW of wavelength $L=20$ at $A=2$. The initial condition was obtained through numerical continuation. We then decreased the bifurcation parameter $A$ at a linear rate $-m$ before instantaneously switching it to $A=\Atarget$ upon hitting the stability boundary. We changed the linear decay rate $-m$ across different simulations and compared the observed time delay $\tdelay$. In other words, we set $A=2 - mt$ for $t<\tstab$, and $A=\Atarget$ for $t>\tstab$, where $\tstab$ denotes the time at which $A=2 - m\tstab = \Astab$. The results, visualised in \cref{fig: suppl: klausmeier memoryless}, highlight that the order of magnitude of the delay $\tdelay$ is unaffected by changes to the model dynamics that occur before crossing a stability boundary in the Busse balloon. We thus term this phenomenon \textit{approximate memorylessness}.

\section{More examples of delay predictions}
\begin{figure}
	\centering
	\includegraphics[height=0.9\textheight]{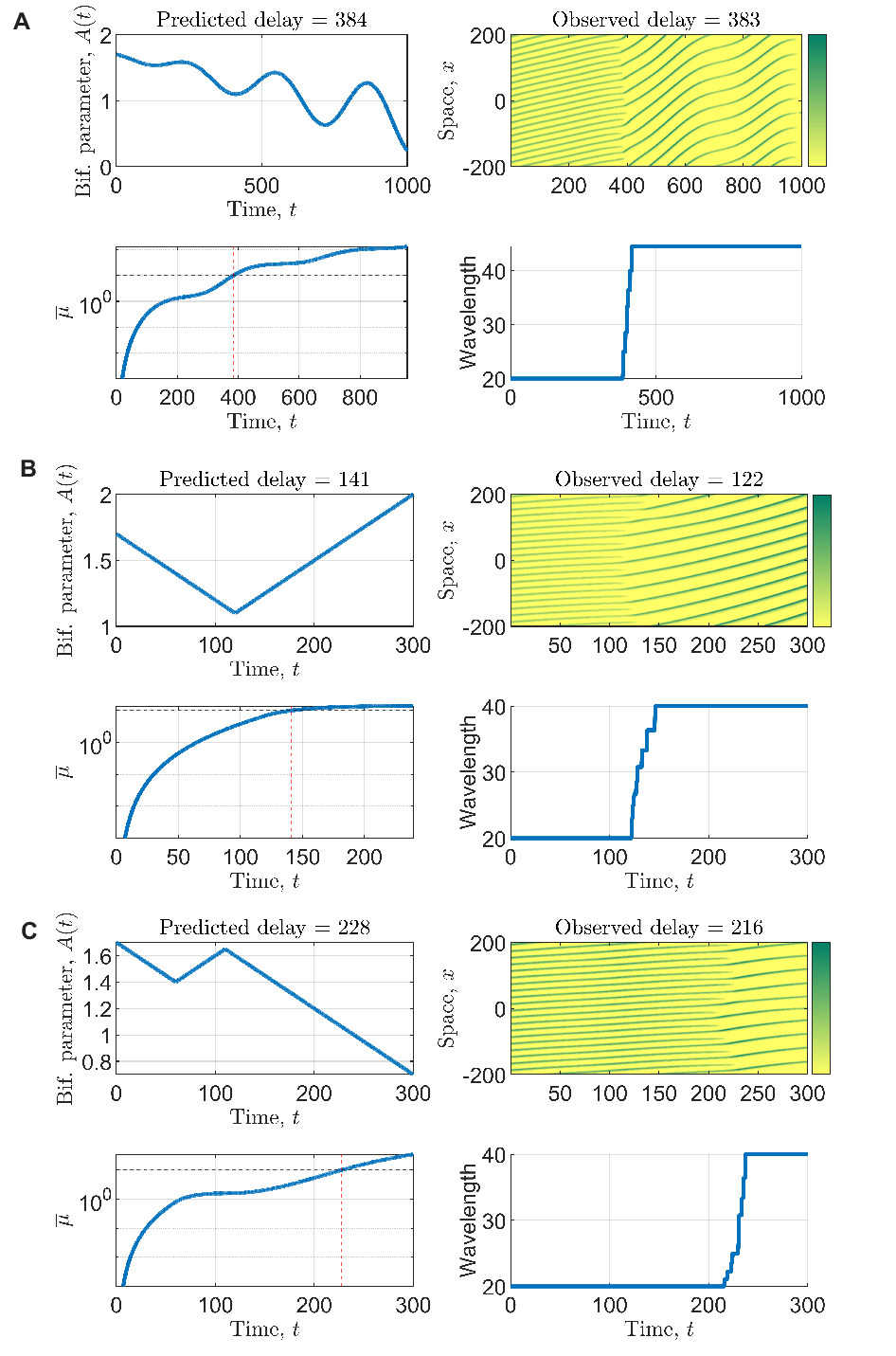}
	\caption{\textbf{More examples of delay predictions in the Klausmeier model.} For a full figure caption, see \cref{fig: delay predictions examples}. For a description of the parameter change regimes used, see the supplemental text.} \label{fig: suppl: more delay predictions}
\end{figure}

In this section, we provide evidence that our method for predicting the time delay between PTW destabilisation and the occurrence of a wavelength change provides accurate order of magnitude estimates not only for the linear parameter change regimes described in the main text, but also for other parameter change regimes. In particular, we tested system \eqref{eq: klausmeier model} in the following cases:
\begin{itemize}
    \item a decaying sinusoidal regime: $A(t) = 1.7 - mt(2+\sin(t/50))$, $m = 0.0005$ (\cref{fig: suppl: more delay predictions}A),
    \item a ``change of direction'' regime: $A(t) = 1.7-mt$ for $t<t^\ast$, $A(t) = 1.7 - mt^\ast + m(t-t^\ast)$ for $t>t^\ast$, $m=0.005$ (\cref{fig: suppl: more delay predictions}B),
    \item a ``zig zag'' regime:  $A(t) = 1.7-mt$ for $t<t^\ast = 60$, $A(t) = 1.7 - mt^\ast + m(t-t^\ast)$ for $t^\ast<t<t^{\ast\ast} = 115$, $A(t) = 1.7 - m(t^{\ast\ast} - t^\ast) - m(t-(t^\ast+t^{\ast\ast}))$ for $t^\ast<t<t^{\ast\ast} = 115$, $m=0.005$ (\cref{fig: suppl: more delay predictions}C).
\end{itemize}
In all cases, we observed excellent agreement between the order of magnitude in the prediction and the numerical simulation.

\section{Results for the mussel model}\label{sec: mussel results}

\subsection{The sediment accumulation model for intertidal mussel beds}\label{sec: mussel model def}
 We further to highlight that our results on the delayed loss of stability of PTWs applies to a range of different models admitting PTWs. We thus also consider the \textit{sediment accumulation model} for intertidal mussel beds by \textcite{Liu2012a,Liu2014}. For full details on the model, we refer to \textcite{Sherratt2016b,Liu2012a,Liu2014}. More information on the ecological phenomenon, as well as an alternative mathematical model, can be found in \textcite{Koppel2005,Shen2020,Bennett2018,Sherratt2016a}. Suitably nondimensionalised (\textcite{Liu2012a}), the model we consider is

\begin{subequations}
		\label{eq: mussel model}
		\begin{align}
            \frac{\partial m }{\partial t} &= \overbrace{\frac{\delta a m(s+\eta)}{s+1}}^{\text{mussel growth}} - \overbrace{ m}^{\text{mussel death}} + \overbrace{\frac{\partial^2 m}{\partial x^2}}^{\text{mussel dispersal}}, \\
            \frac{\partial s }{\partial t} &= \underbrace{m}_{\text{sediment build-up}} -\underbrace{\theta s}_{\text{sediment erosion}} + \underbrace{D\frac{\partial^2 s}{\partial x^2}}_{\text{sediment dispersal}}, \\
            \frac{\partial a }{\partial t} &= \underbrace{1+\varepsilon a}_{\substack{\text{transport from}\\ \text{upper water layers}}} - \underbrace{\frac{\beta a m(s+\eta)}{s+1}}_{\text{algae consumption}}  + \underbrace{\nu\frac{\partial a}{\partial x}}_{\substack{\text{algae flow}\\ \text{with tide}}}.
		\end{align}
	\end{subequations} 
 The densities $m(x,t), s(x,t),$ and $a(x,t)$ describe the mussel density, sediment density, and algae density, respectively at space point $x\in \R$ and time $t\ge0$. The main bifurcation parameter of the system is the mussel growth rate $\delta \ge 0$. The system fits into our general framework \eqref{eq: general model} by setting $u = (m,s,a), \alpha = (\delta, \eta, \theta, \varepsilon, \beta), d = (1,D,0)$ and $f(u;\alpha) = (\delta a m (s+\eta)/(s+1) - m, m-\theta s, 1+\varepsilon a  - \beta a m (s+\eta)/(s+1))$.

\begin{figure}
	\includegraphics[width=\linewidth]{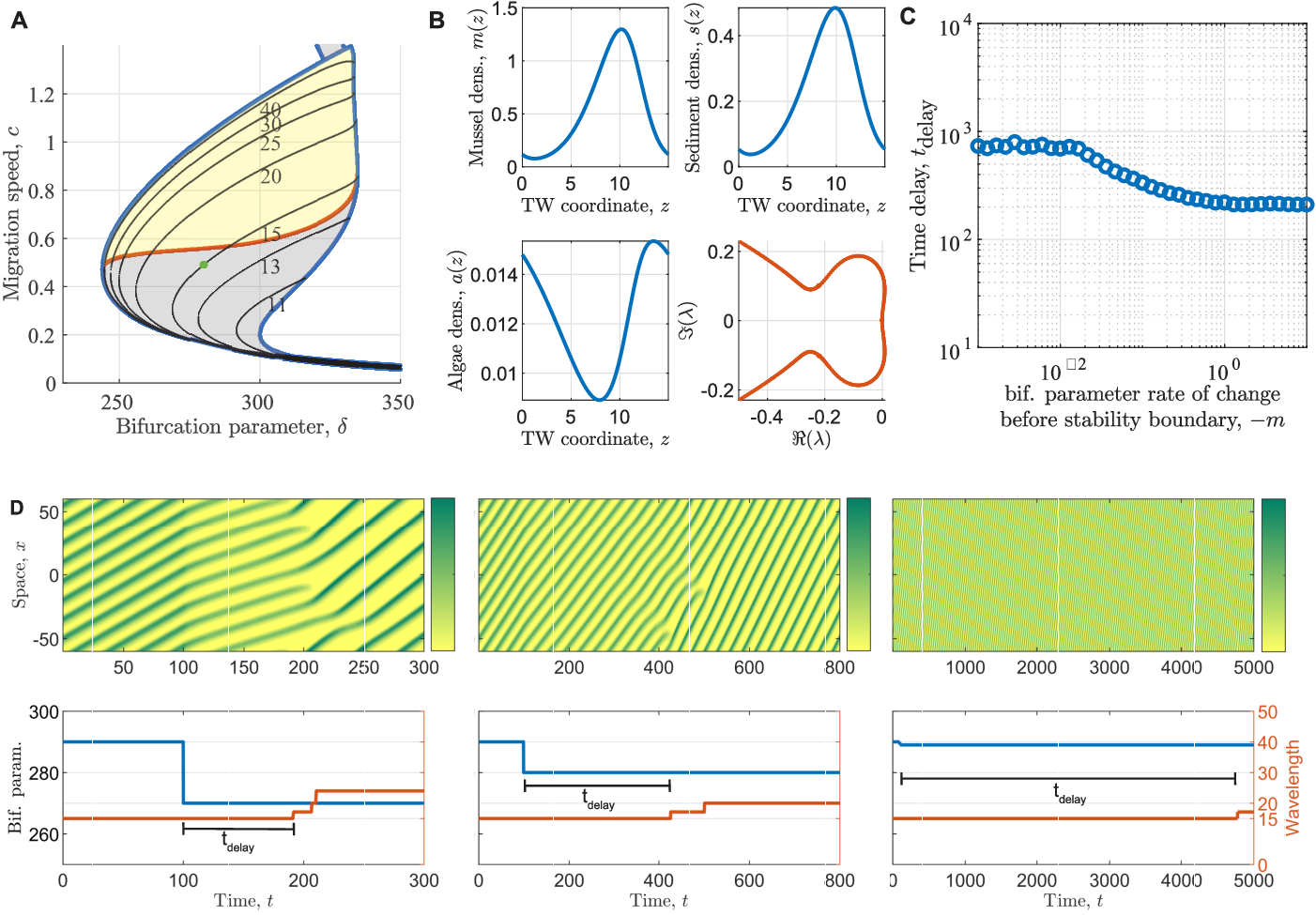}
	\caption{\textbf{Busse balloon, essential spectra, and wavelength changes in the mussel model.} \textbf{A:} Busse balloon of the sediment accumulation model for intertidal mussel beds system \eqref{eq: mussel model}. Shaded regions visualise regions of pattern existence, split into stable (yellow) and unstable (grey) patterns. Existence boundaries are shown in blue, stability boundaries in red. Annotated solid black curves show wavelength contours. The green dot on the $L=15$ contour indicates the location of the solution shown in \textbf{B}. Note that the blue curve in the top right region splitting stable into unstable regions is not a stability boundary. Rather, it is the location of a homoclinic orbit in which the stable solution terminates and leaves only one unstable solution remaining (below the homoclinic orbit one stable and one unstable solution exist; for full details on the Busse balloon for this model see \textcite{Sherratt2016b}  \textbf{B:} One period only of an example PTW for $\delta=280$ and $L=15$ is shown (blue). Its essential spectrum (red) is shown in the bottom right panel. \textbf{C:} The time delays, $\tdelay$, for simulations with varying rate of change ($-m$) of the bifurcation parameter before hitting the stability boundary ($\delta=310 - mt$ for $t<\tstab$), and subsequent instantaneous switch to $\delta=\deltatarget=280$. Note that the delay is approximately independent of $m$ (compared to the order of magnitude differences reported in \textbf{D}). \textbf{D:} Examples of wavelength changes occurring after a time delay. The top panel in each of the rows shows the contour plot of the mussel density in the time-space parameter plane. The simulation is initialised with a stable PTW constructed using numerical continuation. The bifurcation parameter (blue curve in bottom panel) is kept at its initial value $\delta_0 = 290$ for 100 time units before it is abruptly decreased to $\delta=\deltatarget$ beyond the stability boundary, which is located at $\deltastab \approx 299$. Here, $\deltatarget = 270$ (left), $\deltatarget = 280$ (middle), and $\deltatarget = 289$ (right). A wavelength change (red curve in bottom panel) only occurs after a time delay $\tdelay$. Note the different limits on the time axes. Other parameter values are $\varepsilon = 50$, $\beta = 200$, $\eta = 0.1$, $\theta = 2.5$, $\nu = 360$, and $D=1$ across all figures.}\label{fig: suppl: mussels busse spectra wavelength changes}
\end{figure}

\begin{figure}
	\includegraphics[]{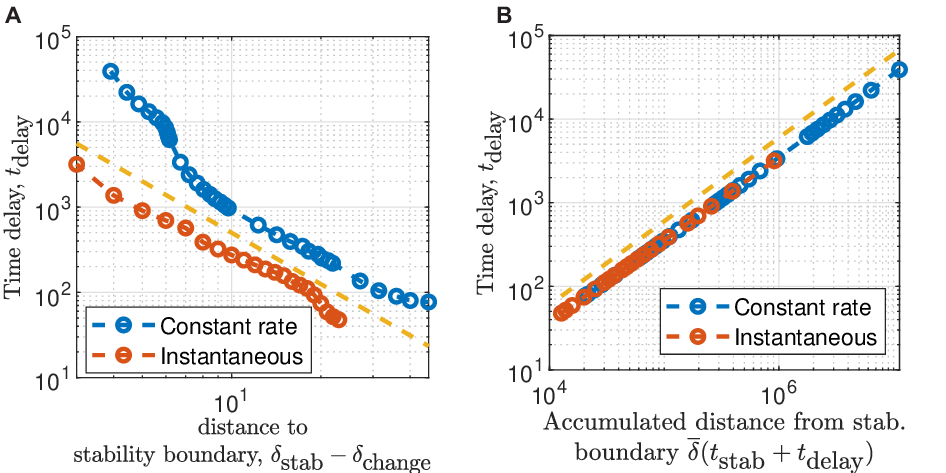}
	\caption{\textbf{Time delay in relation to bifurcation parameter changes.} \textbf{A}: The relation between the time delay $\tdelay$ and the distance of the bifurcation parameter $\delta$ from the stability boundary at the time of the wavelength change is shown for an instantaneous parameter change to a target value $\deltatarget$ (red), and for a regime in which $\delta$ decreases at constant rate $m$ (blue) (where each data point corresponds to a different value of $0.0001\le m \le 0.03$). The dashed lines have slope $-2$. \textbf{B}: The relation between the time delay $\tdelay$ and the accumulated distance from the stability boundary at the time of the wavelength change, $\overline{\delta}(\tstab + \tdelay)$, is shown for both parameter change regimes.  The dashed line has slope $1$. Other parameter values are $\varepsilon = 50$, $\beta = 200$, $\eta = 0.1$, $\theta = 2.5$, $\nu = 360$, and $D=1$}\label{fig: suppl: mussels delay vs delta}
\end{figure}

\begin{figure}
	\includegraphics[]{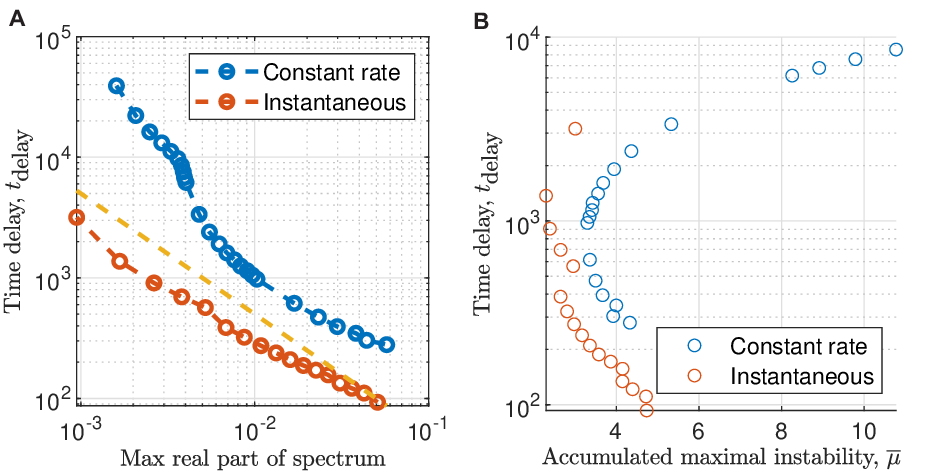}
	\caption{\textbf{Time delay of wavelength change in relation to the maximum real part of the essential spectrum.} \textbf{A}: The time delay of a wavelength change that occurs after crossing a stability boundary is compared with the maximum real part of the essential spectrum of the unstable solution at which the wavelength change occurs for instantaneous changes of the bifurcation parameter (red), and constant rates of change of the bifurcation parameter $m$ (blue) (where each data point corresponds to a different value of $0.0001\le m \le 0.03$). The dashed line has slope -1. \textbf{B}: The time delay is compared with the accumulated maximal instability for both parameter change regimes. Other parameter values are $\varepsilon = 50$, $\beta = 200$, $\eta = 0.1$, $\theta = 2.5$, $\nu = 360$, and $D=1$.}
	\label{fig: suppl: mussels delay vs max real part spectrum}
\end{figure}

We repeated our analysis shown in the main text for the sediment accumulation model describing the formation of intertidal mussel beds, shown in \eqref{eq: mussel model}. Like in the Klausmeier model, PTW of \eqref{eq: mussel model} lose their stability for decreasing bifurcation parameter (here the mussel growth rate $\delta$) at an Eckhaus stability boundary (\cref{fig: suppl: mussels busse spectra wavelength changes}A-B). We found that all results obtained for the Klausmeier model carry over to the mussel model. In short, we found that
\begin{itemize}
    \item there is a delayed loss of stability phenomenon and observed time delays between PTW destabilisations and wavelength changes cover several orders of magnitude (\cref{fig: suppl: mussels busse spectra wavelength changes}D),
    \item the delay is approximately memoryless to dynamics that occur before crossing the stability boundary (\cref{fig: suppl: mussels busse spectra wavelength changes}C),
    \item the relation between the time delay $\tdelay$ and the distance to the stability boundary at the wavelength change $\deltastab - \deltajump$ is $\tdelay \sim (\deltastab - \deltajump)^{-2}$, but there are an order of magnitude differences in $\tdelay$ for different parameter change regimes (\cref{fig: suppl: mussels delay vs delta}A),
    \item the relation between the time delay $\tdelay$ and the accumulated distance from the stability boundary $\overline{\delta}(t)$ at the wavelength change is $\tdelay \sim \overline{\delta}(\tstab + \tdelay)$, with no quantitative differences between different parameter change regimes (\cref{fig: suppl: mussels delay vs delta}B),
    \item the relation between the time delay $\tdelay$ maximum real part of the PTWs' essential spectra $\mu(\delta)$ is $\tdelay \sim \mu(\deltajump)^{-1}$ but there are an order of magnitude differences in $\tdelay$ for different parameter change regimes (\cref{fig: suppl: mussels delay vs max real part spectrum}A),
    \item the accumulated maximal instability at a wavelength change $\mubar(\deltajump) \in I_{\mujump} \approx [3,11]$ consistently for all parameter change regimes (\cref{fig: suppl: mussels delay vs max real part spectrum}B). 
\end{itemize}

\end{document}